\magnification=\magstephalf\hsize=14truecm\hoffset=1truecm\parskip=5pt
\centerline{Consistent-Histories Description of A World with Increasing Entropy}
\vskip .4cm
\centerline{C. H. Woo}
\centerline{Physics Department}
\centerline{University of Maryland}
\centerline{College Park, MD 20742}
\vskip .5cm
\line {1. Attempt at a self-contained quantum description\hfil}

The consistent-histories program$^{1,2,3}$ represents a more concerted effort than   
previous attempts to overcome what some workers regard as the main defect of the
Copenhagen interpretation: the need to invoke the existence
of a classical world as a pre-condition to the meaningful interpretation of quantum mechanics. 
Hence, just as for any other attempt at a self-contained quantum description, the first 
order of business for this program is to represent the classical world correctly.  It is precisely on this point, however, that the consistent-histories program has not yet been carried to completion to the satisfaction of all concerned.  It is not the purpose of
this note to discuss the nature of the difficulties encountered; rather, 
assuming that there is a quantum description of the quasi-classical world
along the lines of consistent-histories, I ask what the description will be like.
In particular, I investigate the rule of association between the histories in a
suitable family and the quasi-classical worlds that the formalism is supposed to describe.

A family of consistent histories
is specified by an initial state $\rho(0)$, the Hamiltonian of the world $H$, and 
sequences of events.  Each sequence is represented by a chain of Heisenberg-picture projectors:
$$C(\alpha_1\alpha_2...\alpha_n)\equiv E^{\alpha_1}_1(t_1)E^{\alpha_2}_2
(t_2)...E^{\alpha_n}_n(t_n)\eqno(1),$$ where the subscript on the projector $E^{\alpha_i}_i$
refers to the nature of the resolution of the identity, and the superscript to the
specific element within that set of projectors. The probability $P(C)$ for the 
occurrence of a history corresponding to the chain $C$ is: $$P(C)=Tr(C^+\rho(0)C)\eqno(2).$$
All histories within a family must fulfill certain consistency conditions so that
the probabilities for their being realized satisfy normal additivity rules.  The
Eq.(1) has been given in the simplest form which does not take the possible 
branch-dependence of projections into account, but I will later return to the issue
of branch-dependence where it might be relevant to the subject of this note.

Consistency alone does not guarantee that the events in a history correspond to
what in the Copenhagen interpretation would be called ``actualized" or ``registered"
events.  In the absence of external observers, it appears that a selection criterion 
needs to be added to the consistent-histories formalism to characterize such special
events, which correspond to the ordinary experience that these
occurrences actually happened.  It turns out that 
a mathematical formulation of such a criterion is no easy task.$^{4,5}$  But the
physical basis that qualifies certain types of events as actualized can be 
examined, and is discussed in some detail in the Appendix.  Roughly speaking, the 
essence of actualization is verifiability from accessible records; and events which may
be regarded as having been actualized or registered will
be called ``verifiable" from here on.  A verifiable event is different from
a robust event$^6$, the consistency of which is maintained only by decoherence.
Decoherence ensures that the different alternatives will not later interfere, but
does not guarantee that only one of the alternatives is ``actually realized."  It is
only when the accessible records in a branch can corroborate which alternative is
realized that one has a verifiable event.  A similar point has been made
by Zurek and co-workers within the framework of what they call ``the existential
interpretation of quantum mechanics,"$^7$ although it is not always clear whether
they want to make a sharp distinction between a robust event, where the ``record"
may consist of little more than scattered photons which escape to infinity, and a verifiable event supported by accessible records.  The main point of this note
is sufficiently simple that the precise mathematical conditions qualifying an
event as verifiable, which have resisted formulation so far, are 
not needed; but it is necessary to assume that a selection criterion 
does exist.  The formulation of a set of clearly stated conditions for 
distinguishing certain events as verifiable is absolutely essential for the completion of the
consistent-histories program, because this program tries to describe the classical
world with events, and the quantum events which are directly relevant to the description of
occurrences in the observable world are exclusively verifiable events. 

Already in classical physics it is imagined that there exists an underlying 
fine-grained structure, the complete details of which can never be checked 
within the limitations imposed by the availability of resources.  Turning to a quantum 
description of the world adds the need to consider superpositions with phase 
correlations, and subtracts the possibility of certain types of simultaneously
precise data, but it is not greatly different in spirit as far as 
entertaining the idea of the world as a closed system is concerned.  In particular, 
among classical statistical physics expositions there are statements like ``the 
entropy of the world, regarded as a closed system, is non-decreasing."  The question 
this article addresses is: What is the nature of the appropriate coarse-grained 
quantum state, so as to describe
a quasi-classical world in which entropy increases?    

It should be made clear at the outset that the goal here is only to clarify the 
nature of the appropriate quantum representation.  There is 
no claim that this quantum description provides any {\it a priori} explanation for 
irreversibility -- the common assumption that the world, under suitable coarse-graining, 
was in a particularly low entropy state 
once, and is still relatively low in entropy today, as the cause for the validity of 
the second law in the present era, will again be needed (in the form of an assumption
on the near saturation of records,
see section 3).  But even with such a modest goal something can be learned.  
Because the aim of a self-contained quantum description of the world is so ambitious, whereas 
the language of consistent histories is so very economical -- thus far the branching of
histories is essentially the only type of events seriously considered -- it 
turns out that even the mere task of describing entropy increase requires some modification of that perspective.  The modification consists of the realization that the process of 
merging two or more histories together is as relevant to describing the observable world
as the opposite process of splitting a history into separate branches.  As a consequence,
the number of possible quasi-classical branches does not increase indefinitely.  In
other words, there is no population explosion of the kind in Everett's many-world picture. 
\vskip .5cm
\line{2. Average entropy is non-increasing under branching\hfil}
\vskip .2cm

Although some workers prefer to speak of only histories and not the instantaneous
density matrix of the world, most researchers, once taking the plunge of considering
a self-contained quantum description of the world as a whole, 
do not seem to object to this notion.
Since the purpose here is to provide a quantum representation of the change in time 
of the macroscopic world of verifiable
facts, it is appropriate to have an instantaneous density matrix to correspond to
the situation at a specific instant.  The consistent-histories formalism
suggests a natural density matrix to consider.
The matrix $\rho(C,t_n)$ defined by
$$\rho(C,t_n)=C^+(\alpha_1...\alpha_n)\rho(0)C(\alpha_1...\alpha_n)/
Tr[C^+(\alpha_1...\alpha_n)\rho(0)C(\alpha_1...\alpha_n)]\eqno(3a)$$
is a suitable candidate, because it has the property that the conditional
probability for the outcome of the next event to be that particular alternative 
represented by $E^{\alpha_{n+1}}_{n+1}(t_{n+1})$, given the fact that the past
history is already specified by $C(\alpha_1...\alpha_n)$, equals appropriately 
$Tr[\rho(C,t_n)E^{\alpha_{n+1}}_{n+1}(t_{n+1})]$.   The density matrix is
alternatively defined by the recursive relation
$$\rho(C(\alpha_1...\alpha_n),t_n)=E^{\alpha_n}_n(t_n)\rho(C(\alpha_1...\alpha_{n-1}),
t_{n-1})E^{\alpha_n}_n(t_n)/N$$
where$$N\equiv Tr[E^{\alpha_n}_n(t_n)\rho(C(\alpha_1...
\alpha_{n-1}),t_{n-1})]\eqno(3b).$$ 
The way this density matrix enters the probability for the next event recommends
it as a candidate for a quantum representation for the instantaneous
state of the world, provided $\rho(0)$ is a suitable initial state of the world.  
 
Only three things enter into $\rho(C,t_n)$: the initial density matrix, the Hamiltonian,
and the chain of projections.  The Hamiltonian is presumably fixed once and for all,
and so is the initial density matrix.  What changes as time progresses is, in the
usual consistent-histories formalism, an indefinite elongation of the 
chain of projections.  For a verifiable history, the projections are 
all supported by classical records, and hence
the increase in the information content of the records in the classical world which $\rho(C,t_n)$ is supposed to describe, due to the occurrence of a new event, is 
no less than the increase in the information content of $\rho(C,t_n)$ itself.  The increase of the information content of classical records can be greater,
however, because for the quantum state some of the initial information
may have been destroyed by the new projection (noncommutivity).   One notes that
the information content of classical records is really no different from that of
the classical world.  Mallarme famously said:``Everything there is in the world
exists to be put in a book," but the ``book" should be generalized to include
all types of classical records.

As to the quantum state, since the choice of an initial state must ultimately be justified by 
empirical support, from the position of trying to obtain a
quantum description of the classical world that stays as close to empirical facts
as possible, there is no good reason to treat the initial state and later events by
different principles.  If the later events are factual only by virtue of the records available today, then the initial state is to be determined by the same criterion: maximum
ignorance consistent with present data.  Indeed, one could let $\rho(0)$ be 
proportional to identity, and put in the earliest known
information through the first projection at time $t_1$, and in this way the initial
data and subsequent events are treated on the same footing.  
Formally, the method of choosing 
a least-biased state by a variation principle subject to constraints is
well known, although in practice this will be incredibly complicated for the
real world.  If one wants, however, to have a quantum description that is as
close to the classical world as possible, leaving out no relevant classical
data but also putting in as little extraneous input as possible,
then it is the maximally detailed verifiable histories with
the least-biased initial state that are relevant.  In other words, in principle
one would like to have every piece of factual data in the classical world
represented in the quantum description: if a classical datum arises out of deterministic 
evolution, the information resides in the initial state and earlier events; 
and if it arises out of an
amplified quantum fluctuation, then the datum emerges at the point 
of a new verifiable event.
The objective of the consistent-histories program to have a self-contained
quantum description of the quasi-classical world is not realized unless there
is a family of such ``least-biased and maximally detailed verifiable histories."
It will be understood from here on that
this is the kind of history which we are concerned with.
 
From the consideration of information gain, one might think that for such a
least-biased and maximally detailed verifiable history the change in the
von Neumann entropy $s[\rho(C,t_n)]\equiv 
-Tr[\rho(C,t_n) log \rho(C,t_n)]$ due to the occurrence of a new verifiable event 
has at least the same sign as the 
change in the statistical entropy of the 
classical world that this history describes.  Making this association, however, 
gives rise to an immediate difficulty.  We state this difficulty as follows:
\vskip .2cm
\line{\bf Proposition\hfil}
\vskip .1cm
Suppose that (i) there is a faithful quantum description of quasi-classical worlds by 
means of a family of least-biased and maximally detailed verifiable 
histories, which undergo only branchings, and (ii) the
change in the von Neumann entropy of the density matrix of Eq.(3) due to 
a verifiable event is in the same direction as the change in the 
classical statistical entropy of the world being described.  Then on the
average the second law is violated in these worlds.
\vskip .2cm 

This proposition follows directly from a theorem of Groenewold and Lindblad.  Groenewold
first conjectured$^8$ that under a branching the average entropy of a branch 
is equal to or less than that of
the parent-state, but a geometrical approach to proving this conjecture turned out
to be difficult.  Lindblad$^9$ then gave an elegant proof drawing on the
some previous results.$^{10}$ 
\vskip .3cm
\line{{\bf Theorem} (Groenewold-Lindblad)\hfil}
$$If\ \rho'_{\alpha}\equiv E^{\alpha}\rho E^{\alpha}/p_{\alpha}, \ and\ p_{\alpha}
\equiv Tr(E^{\alpha}\rho),$$
$$\overline{s[\rho']}\equiv\sum_{\alpha} p_{\alpha}s[\rho'_{\alpha}]\le s[\rho] \eqno(4).$$ 

It is understood here that the set $\{E^{\alpha}\}$ consists of orthogonal projectors
which resolve the identity.  Since any chain of projections entering into the 
specification of a history is built up from projections satisfying the conditions
of this theorem, it follows that, when averaged over the branches of each
splitting, the entropy for the density matrix 
$\rho(C,t_n)$ is non-increasing with $n$, regardless of what kind of coarse-graining
has been built into the projections.  This theorem does not rule out the 
possibility that along a particular branch the entropy is non-decreasing, but that 
would be an exceptional situation requiring an explanation, 
for it goes against the average trend of
Eq.(4), which is for the entropy associated with $\rho(C,t_n)$ not to increase 
with increasing $n$.

\vskip .3cm
\line{3. The merging of histories\hfil}
\vskip .2cm
We have used expressions like ``the observable world" and ``accessible records"  without 
being specific on ``observable" and ``accessible" to whom.  This is because a detailed
examination would have involved a lengthy digression, and I decided to relegate all such
matters to an appendix.
At this point let it be tentatively supposed that these expressions have sensible meanings.  
Then the entropy-decreasing tendency of branching is at least intuitively
comprehensible from the information-gain argument made earlier.  This argument suggests that
entropy-decrease becomes unavoidable for a history in which all the events are verifiable.   The quoted theorem says that entropy tends to decrease even when earlier information can be partially destroyed, but for verifiable
events the records keep the earlier information intact, thus 
making it even more likely that entropy would decrease with the
prolongation of a history, i.e., with the accumulation of more and more records.

The most natural way out of the quandary created by a tendency for the von Neumann entropy to decrease under branching is to take into account the fact that in a world 
of increasing entropy records do decay.  If a branching in the past is still relevant later because of the persistence of its associated accessible records, then the decay of records,
by their becoming inaccessible as evidence,  
would partially undo the effect of that branching.  Such partial undoing is what I call ``merging."  It is desirable to incorporate merging without spoiling
the simplicity of the consistent-histories approach.  The following proposal for 
bringing the deterioration of records into the quantum description 
of a closed system is guided by two considerations:
(a) the verifiability of a given event is a time-dependent property, since records
change with time; and (b) the decay of the records concerning a past event 
is not the same as a quantum erasure.  The implication of the first premise is that,
unlike a branching which involves just one time, a merging commonly involves two times: an
earlier moment when the event and its registration occurred, and the later moment when the records concerning
the outcome of that event are obliterated.  The implication of the second premise is that the
decay of records does not completely undo a projection.  If one were to represent 
the decay of the records concerning an event at time $t_i$ by 
removing the projection $E^{\alpha_i}_i(t_i)$ altogether 
from the chain $C$ which helps to define $\rho(C,t_n)$, it would be as if no event 
happened at $t_i$ at all; whereas the decay 
of records presupposes that an event did occur, and it was even verifiable at one 
stage, and the different outcomes were decoherent.  When the records 
decay at time $t_n$ and it is no longer possible to verify the outcome of what
actually happened at $t_i$, the relative likehood of the various alternatives 
being the best that can be deduced from the surviving evidence, then these alternatives 
are to be incoherently summed.  In contrast, removal of the projection at 
$t_i$ would correspond to a quantum erasure, with the alternative components 
added back together with exactly the correct phases, which is 
imaginable in a laboratory setting but is not realistic for the overwhelming
majority of actual events.  The expression ``merging of histories" for
describing the incoherent summation seems appropriate because of the following
consideration.  The most common pictorial representation for the structure of a family of
histories is that of a branching tree.  Although the summation over those
branches which the records can no longer distinguish is akin to bundling several
branches together rather than fusing them, the subsequent offshoots undergo a real 
reduction in number.  For example, a branching into $n$ branches followed by a second 
splitting each into $m$ branches would result in a total of $nm$ alternatives; 
but if the records for the first branching are later destroyed, the final outcome 
leaves only $m$ alternatives, and in that sense the $nm$ branches have merged 
into the $m$ branches.  The situation described corresponds to a family
structure where the projections are not branch-dependent.  If the later 
projections depend on which branch the event is taking place in, 
ther situation becomes more complicated.  But the
existence of branch-dependence by itself provides some sort of a record, and
therefore in considering the complete destruction of records, we limit ourselves to
situations where branch-dependence is absent.

The mathematical formulation of merging is relatively straightforward.  
Thus the erosion at time $t_n$ of the records concerning an event at $t_i$, where
$t_n>t_i$, is to 
be represented by the transformation $T$:
$$T:  \rho(C(\alpha_1...\alpha_i...\alpha_{n-1}),t_{n-1})\rightarrow \rho
(\bar{C}(\alpha_1...\bar{\alpha}_i...\alpha_{n-1}),t_n)\equiv$$
$$\sum_{\alpha_i}b_{\alpha_1...\alpha_i...\alpha_{n-1}}
e^{-iH(t_n-t_{n-1})}\rho(C(\alpha_1...\alpha_i...\alpha_{n-1}),t_{n-1})
e^{iH(t_n-t_{n-1})}$$
where
$$p_{\alpha_1...\alpha_i...\alpha_{n-1}}\equiv Tr[C^+(\alpha_1...\alpha_i...\alpha_{n-1})
\rho(0)C(\alpha_1...\alpha_i...\alpha_{n-1})]$$
$$b_{\alpha_1...\alpha_i...\alpha_{n-1}}\equiv p_{\alpha_1...\alpha_i...\alpha_{n-1}}/
\sum_{\alpha'_i} p_{\alpha_1...\alpha'_i...\alpha_{n-1}}\eqno(5).$$

This corresponds to a situation where the last step in the destruction of records is
through the deterministic processes which result from the evolution generated by \hfil\break
$exp[-iH(t_n-t_{n-1})]$.  If, on the contrary, the destruction of records is
itself accompanied by the actualization of a new quantum event, then one has instead:
$$T':\rho(C(\alpha_1...\alpha_i...\alpha_{n-1}),t_{n-1})\rightarrow \rho
(\bar{C}(\alpha_1...\bar{\alpha}_i...\alpha_n),t_n)\equiv$$
$$\sum_{\alpha_i}b_{\alpha_1...\alpha_i...\alpha_n}
\rho(C(\alpha_1...\alpha_i...\alpha_n),t_n)$$
where
$$p_{\alpha_1...\alpha_i...\alpha_n}\equiv Tr[C^+(\alpha_1...\alpha_i...\alpha_n)
\rho(0)C(\alpha_1...\alpha_i...\alpha_n)]$$
$$b_{\alpha_1...\alpha_i...\alpha_n}\equiv p_{\alpha_1...\alpha_i...\alpha_n}/
\sum_{\alpha'_i}p_{\alpha_1...\alpha'_i...\alpha_n}\eqno(6).$$

     The proposal is that the instantaneous quantum state suitable
for describing what is happening in the macro-world corresponds to 
a density matrix of the form $\rho(\bar{C}(\alpha_1...\bar{\alpha}_i...\alpha_n),t_n)$ rather than
of the form $\rho(C(\alpha_1...\alpha_i...\alpha_n),t_n)$.  In other words, the 
coarse-graining necessary for describing the observable world cannot be all effected
through the use of suitably coarse projections alone: certain coarse-graining
requires convex summations, as in Eq.(5) and Eq.(6).  The resulting change in the
formalism is very minor, and in particular the transformation corresponding to a
branching is given
by:$$\rho(\bar{C}(\alpha_1...\bar{\alpha}_i...\alpha_{n-1}),t_{n-1})\rightarrow E^{\alpha_n}_n(t_n)
\rho(\bar{C}(\alpha_1...\bar{\alpha}_i...\alpha_{n-1}),t_{n-1})E^{\alpha_n}_n(t_n)/N$$
where $$N\equiv Tr[E^{\alpha_n}_n(t_n)\rho(\bar{C}(\alpha_1...\bar{\alpha}_i...\alpha_{n-1}),t_
{n-1})]\eqno(7),$$
identical in structure to Eq.(3b).  It should be noted, however, that the original
chains $C(\alpha_1...\alpha_n)$ are still relevant because the additional consistency
conditions to be fulfilled with the introduction of any new projection 
$E^{\alpha_n}_n(t_n)$ are in terms of the individual
$C(\alpha_1...\alpha_i...\alpha_n)$ and not in terms of a sum over $\alpha_i$.  

The processes represented by Eq.(5) are clearly entropy non-decreasing.  For the process of Eq.(6) where $t_n>t_i$, or processes where Eqs.(5),(6) and (7) all contribute, there is competition between the entropy non-decreasing
convex summations and the average entropy non-increasing additional branchings.
Without further input it is not possible to say which tendency wins.  If,
however, the world's capability for carrying records is already near saturation,
in the sense that the creation of new records requires in most cases the destruction of some
existing records, then records are almost continuously being created and destroyed.
Over the course of many such creations and destructions there is a sense
in which the entropy-increasing trend wins.  This is because of the fact that
the average decrease in entropy due to
a branching event is necessarily less than or equal in magnitude to the
average increase in entropy when subsequently the records corroborating this
event are destroyed.

This can be proved as follows.  
With the notation of Eq.(4), the absolute magnitude of the average decrease
in entropy as a result of a branching is $s[\rho]-\sum p_{\alpha}s[\rho'_
{\alpha}],$ and the average entropy increase accompanying the destruction of
those records which verify the outcome of the branching is $s[\sum p_{\alpha}\rho'_{\alpha}]-
\sum p_{\alpha}s[\rho'_{\alpha}]$.  The fact that the entropy is non-decreasing
as a result of these two steps follows trivially since 
$s[\sum p_{\alpha}\rho'_{\alpha}]\ge s[\rho]$.

Thus, analogous to the usual assumption that the world is currently in a 
low-entropy state, the operative assumption here that can lead to a
tendency for entropy increase is that the present world is already near saturation for 
record-keeping, so that records are continuously being generated and destroyed.  
The degree of plausibility of this assumption is also 
discussed in the Appendix.

Bennett and Landauer$^{11,12}$ have already pointed out that
entropy increase is associated with the erasure of records.  Their
analysis was in a classical setting, and therefore they
did not address the entropy-decreasing tendency associated with the branching
of quantum histories, nor 
need they be concerned with the difference between classical erasure and quantum 
erasure.  Their starting point involving Maxwell's demons and computers seemingly has
very little in common with our starting point of formulating consistent histories
in order to describe an entropy-non-decreasing world.  Nevertheless their
analysis is relevant to our consideration in two respects.  One is Bennett's
demonstration that it makes sense to speak of the entropy of a single system
instead of that of an ensemble -- the quantum history description of the
quasi-classical world is most simply viewed as dealing with a single system.
The second relevant point is their analysis showing that measurement and copy-making
need not be accompanied by an entropy increase.  Such reversible record-keeping
is highly idealized, and most practical record-making is irreversible.  Nonetheless
as a matter of principle it is important to recognize that record-making is not
unavoidably entropy-increasing, whereas record-erasing is.  Our application of
the average entropy-decrease of Eq.(4) to an amplified quantum fluctuation,
i.e., to a measurement-like event, should be viewed as referring to a measurement
with ideal record-making.  Once the nature of entropy increase is clarified in such ideal
events, then the additional entropy-increase associated with the difference between
ideal record-making and realistic record-making can be accounted for as an accumulation
of elementary events, each accompanied by the destruction of some previously existing records.

There is an alternative possibility for accommodating entropy increase as long
as one is willing to add convex sums to the consistent-histories formalism.  Noting
the existence of many physical
processes having outcomes that are decoherent but not verifiable, one
may be tempted to adopt the rule that every
time such an event occurs the state 
representing the quasiclassical world becomes an incoherent sum over these
alternative outcomes.  This modification will also bring about an entropy
increase.  The defect of this approach is the arbitrariness in the choice of
``relevant variables" which decohere as a result of summing over the ``irrelevant"
variables, that is, an arbitrariness in the division between the environment and
the subsystem of interest.  If the alternatives involving the subsystem are not accessible to verification, why are they ``relevant" and ``of interest"?  Remember that one is attempting a
closed-system description here, and there is no pre-determined environment.  
Our proposal hews close
to what are verifiable, and requires incoherent sums only when there
is a destruction of records, i.e., when formerly verifiable alternatives
later become unverifiable.  

Lastly, although I have argued that the dilemma posed by the Proposition is to be
resolved by adding merging to branching, others
may wish to avoid the conundrum altogether by 
arguing that the change of von Neumann entropy is unrelated to that of the classical
entropy.$^{13}$  If one were to look at the entropy changes
during a measurement-like event, whether it occurred in nature or in the laboratory, 
by the usual way of reckoning, it would indeed appear that the latter position is obviously
valid.  For example, a quantum event having two possible outcomes in a particular
measurement situation can bring about a decrease in the von Neumann entropy of the
density matrix for the quantum system by at most $klog2$, but the recording of the
outcome is usually associated with irreversible processes, causing a rise in the classical
entropy of the world by an amount far exceeding $klog2$.  Thus the signs are opposite
and the magnitudes do not match.  One must remember, however,
from the analysis by Bennett and Landauer that irreversibility is ultimately 
attributable to erasure, that is to say, to processes described by Eq.(5) or
Eq.(6).  Hence the overall change in the statistical entropy of the world can be
accounted for by the change in the von Neumann entropy of $\rho(\bar{C})$.
In other words, the merging of histories has to be taken
into account.  The two changes may not be exactly equal because of the possibility of the
destruction of some quantum information by a new projection, discussed earlier, but
the inequality is such that an increase in classical statistical entropy requires a corresponding increase in the von Neumann entropy of
$\rho(\bar{C})$ if the quantum representation is as close to a detailed
description of the quasi-classical world as possible.  On the other hand, if one insists on having only branching and no merging,
then the change in the von Neumann entropy of $\rho(C)$ is indeed unrelated to the change
in the classical entropy of the world.  But to adopt this as the solution is to unnecessarily
impoverish the consistent-histories formalism, rendering only a very partial quantum description of some aspects of the quasi-classical world.  Such a skeletal description is likely
to be insufficient as replacement for the classical underpinning of the Copenhagen
interpretation.  
     
In summary, the tendency for the von Neumann entropy to decrease under branching, on the average, is a fact to be reckoned with.  This is regardless of how a coarse-grained quantum
state is defined: as long as it undergoes only branchings, the average entropy
tends to decrease with time.  If the second law in the macro-world is taken as 
an input, and if a coarse-grained quantum state is to faithfully describe that
macro-world so that changes of the von Neumann entropy of the quantum state track
those of the classical entropy of the world, then its time evolution has to incorporate the merging of previously verifiably distinct histories.  If the consistent-histories program is to
improve on the Copenhagen interpretation, it has to provide a faithful description of the
quasi-classical world strictly within a self-contained quantum theory.  But that
task is not finished even when a specific family of histories is singled out
by some criterion as being suitable for describing
quasi-classical physics, for there must
be in addition a rule of association between the histories in that family and
a quasi-classical world with all its macroscopic details.  And our conclusion is that even if at one time a single history in this fixed family describes that world, 
it will be the incoherent sum of several histories that has to be associated with the 
same world at a later time.  Another way of saying this is that it is inadequate to use projections alone to represent all coarse-graining: 
there are events which require incoherent sums, besides coarse projectors, to represent 
what is happening to the 
quantum state.  The suggested modifications are easy to incorporate and
does not require any major overhaul of the consistent-histories formalism;
but conceptually the picture of an ever multiplying number of potential
quasi-classical worlds is changed to one where the population of 
quasi-classical worlds need not inexorably grow.
\vskip .3cm

(This paper was presented at the E.H. Wichmann Symposium at the University
of California at Berkeley in June 1999.)    
\vfill\eject

\centerline{\bf Appendix}
\vskip .4cm 

This article contains expressions like ``observable world,"``information,"
``verifiability," and ``accessible records," which inevitably invite questions: Whose 
information?  Observable and verifiable by whom and accessible to whom?
How stable does a physical correlation have to be in order to count as a record?    
Discussion of these issues leads to what many physicists would dismiss as 
``philosophy," and yet avoidance of these issues does not save time.  As 
continuing controversies surrounding the subject of quantum measurement$^{14}$ show, 
some knotty issues refuse to go away.  Even when the final answers are not
available, it is better to state one's tentative understanding as clearly 
and explicitly as possible.

One way to specify the meaning of such expressions is by linking them to 
operations.  The basic issues as far as this paper is concerned
are first of all what kind of an event can be said to have
verifiably occurred, and subsequently when the event can be said to cease to
be verifiable.  In an operational sense, an event is verifiable if the
community of scientists, on the basis of records, 
is capable of reaching a consensus on the outcome, provided efforts are devoted 
to the task of checking this event, limited only by the availability of natural resources.  
Similarly the information content of the
observable world is interpreted to be the total database needed for a complete description 
of the macroscopic world, including every bit of datum that is verifiable.  
This way of describing the relation between records and
the verifiability of an event does not require the records to be unchanging: they
can be evolving in time, because all that
is required is for the scientists to be able to use them to unequivocally
interpret an event in the past.  By the same token, the decay of records can
take many forms: some due to the corruption of macroscopic information through classical
processes, and some through random events
in which quantum fluctuations play a role.  That is why in the sentence following 
Eq.(5) a distinction is made between the decay of records through 
deterministic processes and through quantum fluctuations.  That records can
decay even when the process results from $exp[-iH(t_n-t_{n-1})]$ is not
a contradiction, because realistically the scientific community cannot
completely evaluate this operator within the limits of finite resources,
to the precision needed to overcome widespread deterministic chaos. 

Note that it is potential verifiability that matters rather than actual 
verification: because rigorous verification is costly in 
labor and resources, a coarse-grain description in which 
every datum is actually verified, rather than just being verifiable, would turn
out to be very crude indeed.  In contrast, a description in terms of verifiable
events can be much finer.  The distinction between verified and
verifiable events also helps to answer the following question: With the notion
of information being so closely tied to what the scientific community can verify,
how is the increase in scientific knowledge over time to be reconciled with 
entropy increase?  The answer is that whereas the amount of scientifically 
verified data is increasing, the (much greater) amount of scientifically 
verifiable data is non-increasing when entropy is non-decreasing.   But the 
idea of calling an event ``verifiable" provided only 
that it can be actually verified when attention 
is focused on the task of its validation may seem to lead to a contradiction: the 
attention of the scientific community can after all be turned to checking the
position of a particle with great accuracy or the momentum of a particle with
great accuracy; hence would not the position and momentum be simultaneously
verifiable?  It must be remembered, however, that the branching structure of
a consistent family is regarded as being first given, and the verifiability of
the events entering a history is being considered in that context.  The
consistency conditions exclude histories with events which refer to the
position and the momentum of a particle at the same time.  As long as the
focus of validation is directed to one or another of the events already
in consistent histories, no contradiction would arise.  

By using potential verifiability rather than actual verification 
as a selection criterion, one can envision a relatively refined coarse-grain 
description of the observable world.  In cosmological terms such a stage was reached
probably only after the recombination era: the entropy non-increasing tendency of 
branching discussed in this article becomes relevant only after it is possible 
for branches to be sharply distinguished through the existence of records.  
Initially this tends to bring most branches towards low entropy states.  It 
is only after a vast number of verifiable branchings already left a wealth of 
records, allowing a fairly detailed description of the macro-world,
that new records are mostly made at the expense of 
erasing old ones.   Even then, if an event has a huge number 
of redundant records, the erasing of a few of these will not destroy the
credibility of the event; and therefore the hypothesis of ``near saturation
of record-keeping capacity" presupposes that overall there are far more  
nonredundant records than redundant ones.  In a refined description this is not 
implausible.  The most obvious kinds of redundant records occur right here on
earth, but biologically related individual organisms carry a great deal more
data than their shared genetic information, and redundant records made by
people also do not approach anything close to the capacity of refined
classical information that the earth can carry.

     One may ask why ``scientists" are not replaced by the IGUSes (information 
gathering and utilizing systems) of Gell-Mann and Hartle.  The answer is that 
the greater generality of IGUSes is not particularly helpful in this case: 
an ant is an IGUS, and presumably it has some notion of 
whether some kinds of events happened or not, but one would be most
reluctant to add to the list of verifiable events something that only ants 
are aware of with no possibility of independent checks by human scientists,
now or in the future, even when attention is directed to that event.  In other
words, even with the introduction of IGUSes, reference to the ``community of scientists"
would still be necessary.
    
With the above explanation of what is meant by ``verifiable through records," it is finally
possible to specify, in principle, when a formerly verifiable event ceases to
be verifiable: a verifiable event at $t_i$ may be said to have become unverifiable 
at time $t_n$ if the records in none of the future extensions
from $t_n$ onward would permit
the scientists in those branches to corroborate the outcome of this event.

Although a scientific-community-based approach results in a 
verifiability criterion that is close to the common-sensical notion of 
``objective reality," there are nonetheless some counter-intuitive consequences.  By 
the expression ``scientific community" one usually includes not only today's
scientists but also scientists of the future, because technology improves with
time and defining verifiability as what current technology can
ascertain is too restrictive.  Furthermore, there were no scientists in the 
early stages of cosmological evolution, and yet some of those early occurrences
are regarded as verifiable today, because scientists living considerably later 
than these events can still check them out from the records.  Once it is accepted 
that future scientists must be included in
the notion of a scientific community, one has to face the awkward fact that 
there are different future branches.  Suppose, for example, that a continuation
of our history into two different future branches, branch A and branch B, results
in different consequences for life.  Whereas in 
branch A life continues to thrive and instruments continue to improve,
in branch B life is extinguished forever after a cataclysmic event.  Then according to
the scientific-community-based criterion, there may be current events which 
count as verifiable if our future is in branch A but  
not verifiable if our future is in branch B.
Although from a pragmatic standpoint this difference is immaterial
because present-day scientists cannot check these subtle events in either case,
this example shows the value of an alternative, 
strictly objective standard 
that is independent of the existence of people.  For example, one may try to abstract the
essence of ``verifiability by a scientific community" into a mathematical
criterion for factuality that can be applied 
even when life does not exist.  This objective has not yet been accomplished,
and it is not obvious that it can be reached at all; 
although a formulation in terms of suitable complexity measures, in effect 
having finite-resource computers filling in for scientists, appears promising,
that approach suffers from the defect that some of the notions, such as minimal
program length, though well defined cannot be operationally checked exactly.
There can be heuristic checks, but heuristic standards are likely to be
again evolving with time. 
As to conceptual rigor, separate from the question of mathematical rigor,
one pitfall that is easy to fall into is to unconsciously slip in some criteria which are
reasonable only because of our experience up to now, and then to regard the result as 
being more general than it really is.  For example, the environment-induced superselection
approach implicitly regards cuts between environment and system in ways that are based on our
usual notion of what variables are essentially disentangled from the rest
of the world, and the de Broglie-Bohm theory contains arbitrariness in
giving spatial positions a privileged role.  These arbitrary inputs all seem very
reasonable on the basis of our experience up to now, but then we cannot yet claim
that these theories are totally free from biases associated with people
and their state of advancement.  

In a way it is preferable to explicitly acknowledge this possible lack of finality.  
The dependence on the existence of a scientific community in the
proposed verifiability criterion is similar to the need to refer to normal people 
for Locke's secondary qualities.  Locke's definition is useful only if our
standard for normal people is not going to change significantly in the
future.  Similarly if the standard of objectivity achieved by contemporary
science is not going to undergo substantial improvements in whichever
future branch our world evolves into, then the proposed criterion will be
useful.  Otherwise the utility of this criterion will be limited because of its 
still evolving nature.

\vfill\eject

\line{\bf Notes and References\hfil}
\vskip .3cm
\item{1.} R.B.Griffiths, J.Stat.Phys., {\bf 36}, 219 (1984).

\item{2.} R.Omnes, {\it The Interpretation of Quantum Mechanics}, Princeton
Univ. Press, Princeton (1994); {\it Understanding Quantum Mechanics}, Princeton
Univ. Press, Pinceton (1999).

\item{3.} M.Gell-Mann and J.B.Hartle in {\it Complexity, Entropy, and the
Physics of Information}, W. Zurek ed., Addison-Wesley, Reading (1990).

\item{4.} For example, Hartle and Gell-Mann have presented various
sets of criteria for classicity, but in a telephone conversation with Jim Hartle
in 1998 he informed me that none of these fully satisfied them.  For other
views on some of the difficulties, see, e.g., the reference in note 5.

\item{5.} F.Dowker and A.Kent, J. Stat. Phys. {\bf 82}, 1575 (1996).

\item{6.} The distinction between robust histories and verifiable histories
has been emphasized by the author in Found. of Phys. Lett., {\bf 6}, 275 (1993).

\item{7.} W.H.Zurek, Phil Trans. R. Soc. Lond. A {\bf 356}, 1793 (1998).

\item{8.} H.J.Groenewald, Int. J. Theor. Phys., {\bf 4}, 327 (1971).

\item{9.} G.Lindblad, Comm. Math. Phys., {\bf 28}, 245 (1972).

\item{10.} Readers who are interested in a quick grasp of the
entropy non-increasing nature of branching may consider the
quadratic entropy $s'[\rho]=-Tr\rho^2$ instead of the von Neumann entropy, for
in that case the analog of the Lanford-Robinson result used by Lindblad,
viz., $s'[A+B]\le s'[A]+s'[B]$ for positive trace-class operators $A$ and $B$, is
trivially true.
 
\item{11.} R.Landauer, Rev. IBM J. Res. Dev., {\bf 3}, 183 (1961).

\item{12.} C.H.Bennett, Sci. Am., {\bf 257}, 11-108 (1987).

\item{13.} I thank Lawrence Landau for raising such an objection during
my oral presentation at the Symposium; the following elaboration was 
added subsequently.

\item{14.} For example, see the exchange between J.S. Bell and N.G. van Kampen,
both long immersed in the subject, in 
Physics World, August 1990, p.33, and October 1990, p.20.  

\end